\documentclass[aps,twocolumn,superscriptaddress,prl]{revtex4-1}
\usepackage{pdfpages}
\usepackage{amssymb,amsmath}
\usepackage{graphicx}
\usepackage{hyperref}

\makeatletter
\AtBeginDocument{\let\LS@rot\@undefined}
\makeatother

\begin{document}

\title{Machine Learning Topological Invariants with Neural Networks}

\author{Pengfei Zhang}
\affiliation{Institute for Advanced Study, Tsinghua University, Beijing, 100084, China}

\author{Huitao Shen}
\email{huitao@mit.edu}
\affiliation{Department of Physics, Massachusetts Institute of Technology, Cambridge, Massachusetts 02139, USA}

\author{Hui Zhai}
\email{hzhai@mail.tsinghua.edu.cn}
\affiliation{Institute for Advanced Study, Tsinghua University, Beijing, 100084, China}
\affiliation{Collaborative Innovation Center of Quantum Matter, Beijing, 100084, China}

\begin{abstract}
In this Letter we supervisedly train neural networks to distinguish different topological phases in the context of topological band insulators. After training with Hamiltonians of one-dimensional insulators with chiral symmetry, the neural network can predict their topological winding numbers with nearly 100\% accuracy, even for Hamiltonians with larger winding numbers that are not included in the training data. These results show a remarkable success that the neural network can capture the global and nonlinear topological features of quantum phases from local inputs. By opening up the neural network, we confirm that the network does learn the discrete version of the winding number formula. We also make a couple of remarks regarding the role of the symmetry and the opposite effect of regularization techniques when applying machine learning to physical systems. 
\end{abstract}

\maketitle

Recently, machine learning has emerged as a novel tool for studying physical systems and has demonstrated its ability in problems such as inferring numerical solutions \cite{Arsenault2014,Arsenault2015,Mills2017a,Mills2017}, classifying phases \cite{Carrasquilla2017,Broecker2017a,Chng2016,Zhang2017a,Zhang2017,doi:10.7566/JPSJ.85.123706,doi:10.7566/JPSJ.86.044708,Schindler2017,Ponte2017,Wang2016a,Tanaka2017,VanNieuwenburg2017,Liu2017c,Wetzel2017,Wetzel2017a,Hu2017a,Costa2017,Wang2017,Broecker2017,Chng2017}, accelerating Monte Carlo algorithms \cite{Liu2017,Liu2017a,Xu2017a,Nagai2017,Huang2017a,Huang2017}, detecting entanglement \cite{Lu2017}, and controlling quantum dynamics \cite{6628013,Palittapongarnpim2017,Bukov2017,Dunjko2017}. Among all these applications, learning phases is a particularly intriguing one, as it paves a new route toward discovering new phases or even new physics without prior human knowledge \cite{Schmidt2009}. Indeed, there have already been quite a few works on this direction of identifying phase transitions or even extracting order parameters unsupervisedly, i.e., without the awareness of any concept of phases \cite{Wang2016a,Tanaka2017,VanNieuwenburg2017,Wetzel2017,Hu2017a,Costa2017,Wetzel2017a,Wang2017,Liu2017c,Broecker2017,Chng2017}. 

Aside from the current success of machine learning phases within Landau's paradigm, topological phases are especially challenging to learn for several reasons. First, these phases are characterized by topological properties, e.g. the topological invariants, which are intrinsically nonlocal. Second, these topological invariants are nonlinear with respect to the field configuration. Third, topological invariants are intensive instead of extensive compared to the conventional order parameters. As a result, many commonly used techniques in machine learning turn out to be ineffective. For example, the intensiveness makes it futile to distinguish topological phases with the method of (kernel) principal component analysis \cite{Wang2017}. 

The neural network is nonetheless a promising tool for learning topological phases due to its great expressibility and versatility. Mathematically, these networks are able to approximate any continuous functions if the number of fitting parameters can grow indefinitely \cite{Cybenko1989,Hornik1991}. This great expressibility, together with the development of many effective training algorithms \cite{Glorot2010,Bengio2012,Kingma2014,Srivastava2014,Ioffe2015}, makes the neural network an indispensable ingredient in the boom of modern machine learning \cite{LeCun2015}. In this Letter, we report that properly designed neural networks can successfully learn topological invariants for topological band insulators \cite{Kane2005,Moore2007,Schnyder2009,Kitaev2009a,Chiu2016}. Our formalism and results possess the following key features that make them significantly beyond those in the existing works on this topic: i) The input data are completely local; ii) our study is not restricted to any specific model in the symmetry class; iii) our neural network has generalization power after training. We will elaborate these points in the following. 

To be concrete, we consider one-dimensional topological band insulators of the $ \mathrm{AIII} $ symmetry class \cite{Schnyder2009,Kitaev2009a,Chiu2016}. The general form of such two-band Hamiltonians is $H(k)=\mathbf{h}(k)\cdot \boldsymbol{\sigma}$, where $ \boldsymbol{\sigma}\equiv(\sigma_x,\sigma_y, \sigma_z) $ is the vector of Pauli matrices. The chiral symmetry in $ \mathrm{AIII} $ class requires $SH(k)S^{-1}=-H(k)$. Without loss of generality, we can always choose $ S=\sigma_z $ so that only $h_x$ and $h_y$ are nonzero. In our study, we feed neural networks directly with normalized Hamiltonians $\tilde{H}(k)\equiv \tilde{h}_x(k)\sigma_x+\tilde{h}_y(k)\sigma_y$ at $L$ points discretized uniformly along the Brillouin zone. Here $ \tilde{h}_i(k)\equiv h_i(k)/|\mathbf{h}(k)| $, $ i=x,y $. In other words, the input data are $ (L+1)\times 2 $ matrices of the form: 
\begin{equation}
\begin{pmatrix}
\tilde{h}_x(0) & \tilde{h}_x(2\pi/L) & \tilde{h}_x(4\pi/L) & \ldots & \tilde{h}_x(2\pi) \\
\tilde{h}_y(0) & \tilde{h}_y(2\pi/L) & \tilde{h}_y(4\pi/L) & \ldots & \tilde{h}_y(2\pi)
\end{pmatrix}^T.
\label{eq:data}
\end{equation}
In reality, these input data are available in quantum simulators \cite{Flaschner2016}. In the following, we choose $L=32$ and confirm all our results are insensitive to $L$ as long as $ L \geq 32$.

The topological invariant for the $ \mathrm{AIII} $ class is the winding number, as the first homotopy group of a circle $\pi_1(S^1)$. It is defined for a continuous mapping $ S^1\to S^1: k\mapsto U(k) $, $ k\in[0,2\pi] $. $ |U(k)|=1 $ and $ U(k+2\pi)=U(k) $. For the Hamiltonians given above, we identify $U(k)=\tilde{h}_x(k)+i \tilde{h}_y(k)$. Intuitively, the winding number $ w\in\mathbb{Z} $ is an integer that counts how many times $ U(k) $ winds around the origin when $ k $ changes from $ 0 $ to $ 2\pi $. Its sign denotes the clockwise ($ w<0 $) or the anticlockwise ($ w>0 $) winding. The winding number could be formally computed as
\begin{equation}
w=-\frac{i}{2\pi}\oint_0^{2\pi}  U^*(k)\partial_kU(k)dk. \label{eq:w}
\end{equation}
For discretized $ U(k) $, this reduces to
\begin{equation}
w=-\frac{1}{2\pi}\sum_{n=1}^{L}\Delta\theta(n).
\label{eq:disw}
\end{equation}
where $ \Delta\theta(n)\equiv\left[\theta(n)-\theta(n-1)\right] \mod 2\pi$ so that $ \Delta\theta(n)\in[-\pi,\pi) $ and $\theta(n)\equiv \arg[U(2\pi n/L)]$. 

\begin{figure}[tbp]
	\includegraphics[width=.95\columnwidth]{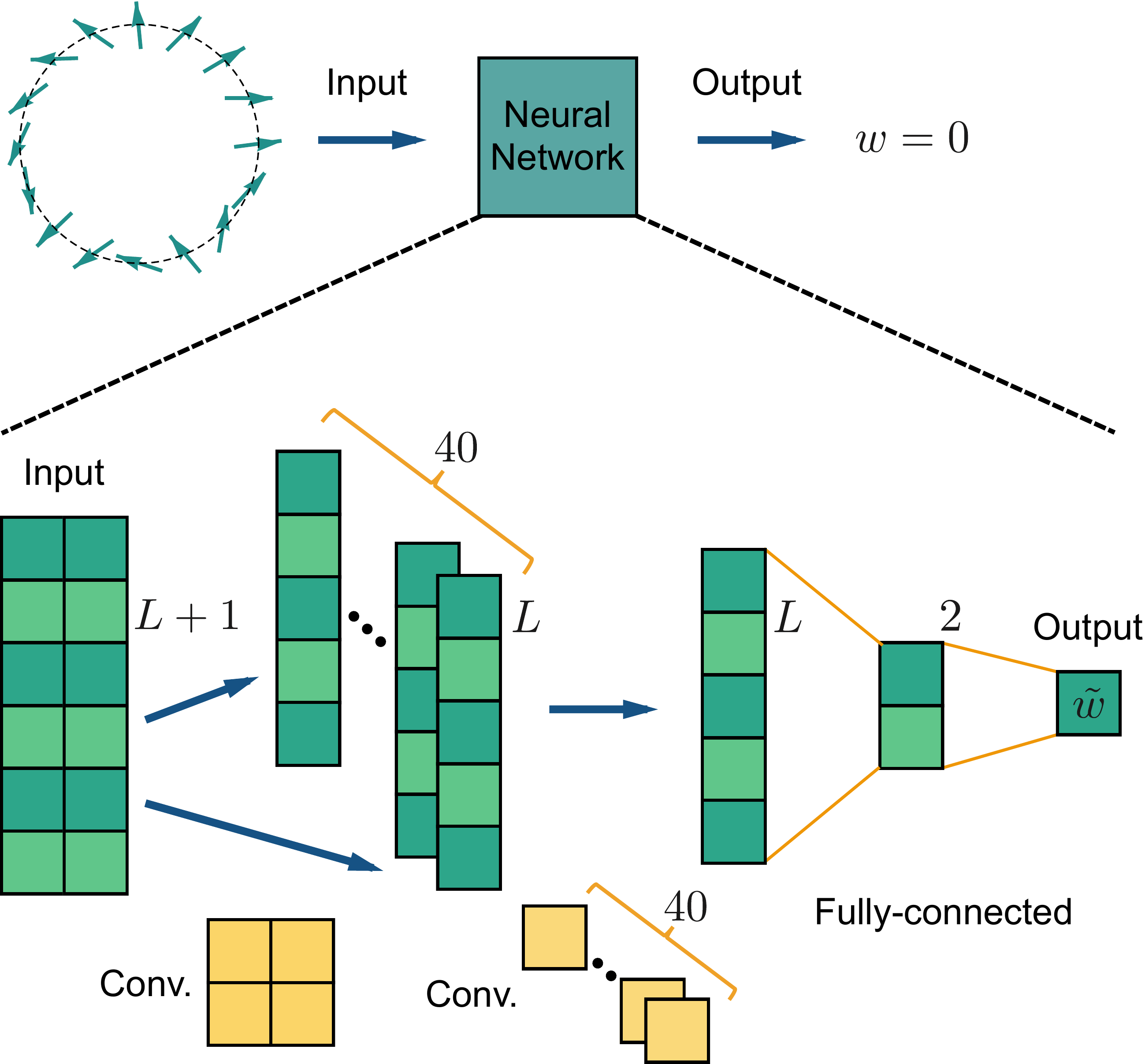}
	\caption{Schematic of the machine learning workflow and the structure of the convolutional neural network. The input Hamiltonians are represented by vectors $ \tilde{\mathbf{h}}(k)\equiv(\tilde{h}_x(k), \tilde{h}_y(k)) $, where $ k\in[0,2\pi] $ is in the Brillouin zone. }
	\label{fig:structure}
\end{figure}

Our machine learning workflow is shown schematically in Fig.~\ref{fig:structure}. The output of the neural network is a real number $ \tilde{w} $, and the predicted winding number is interpreted as the integer that is closest to $ \tilde{w} $. Notice that the input data of form Eq.~\eqref{eq:data} is completely local and highly nonlinear with respect to the formula Eq.~\eqref{eq:disw}. We first train neural networks with both Hamiltonians and their corresponding winding numbers. 
At the testing stage, we feed only the Hamiltonians to the neural networks and compare their predictions with the winding numbers computed by Eq.~\eqref{eq:disw}, from which we determine the percentage of the correct predictions as accuracy. The details of the networks and the training can be found in the Supplemental Material.

The Su-Schrieffer-Heeger (SSH) model  \cite{Su1979} is one of the most simple and widely studied models within the $ \mathrm{AIII} $ symmetry class, whose Hamiltonian is
\begin{equation}
H_{\rm SSH}(k)=(t+t' \cos k)\sigma_x + (t'\sin k) \sigma_y. 
\label{eq:ssh}
\end{equation}
This model hosts two topologically distinct gapped phases with winding number $w=0$ for $ t>t' $ and $w=1$ for $ t<t' $, respectively. We first report the results when the training data are restricted within this model. 

The training set consists of $10^5$ SSH Hamiltonians whose $ (t-t')/t $ are uniformly distributed within $ [-10,10] $, and the test set consists of $ 10^4 $ similar Hamiltonians that are not included in the training set. Surprisingly, even the most simple neural network with no hidden layer nor nonlinear activation function---essentially a linear model used for linear regression---can correctly compute the winding number with nearly 100\% accuracy in the test set after only one training epoch. Further increasing the network complexity by introducing a hidden layer will push the accuracy to exactly 100\%. However, if we test these networks with more general Hamiltonians of winding number $ w=0,1 $, the accuracy sharply drops to around 50\%, which is just the accuracy of blind guesses. This situation could not be improved by increasing model complexity or using more sophisticated neural networks.

Obviously, these networks compute the winding number with a shortcut that is dedicated to SSH Hamiltonians and is only linear with respect to the input data. In fact, due to the additional inversion symmetry in the SSH model $H_{\rm SSH}(k)=\sigma_xH_{\rm SSH}(-k)\sigma_x$, one can read out the winding number directly from the Hamiltonian at the high symmetry point $k=\pi$: 
\begin{equation}
\begin{split}
w=0 & \leftrightarrow \tilde{{\bf h}}(\pi)=(1,0), \\
w=1 & \leftrightarrow  \tilde{{\bf h}}(\pi)=(-1,0).
\end{split}
\label{eq:pi}
\end{equation} 
This local feature is exactly what the networks exploited, for they can predict the winding number perfectly even for $ L=2 $, where only $ \mathbf{h}'(0) $ and $ \mathbf{h}'(\pi) $ are present. 

The lesson is that, if the training data are restricted to some certain model, the neural network would only exploit less universal features of this specific model instead of the universal ones. In the above example, the neural networks do not learn the general formula Eq.~\eqref{eq:disw}, but ``cleverly'' reduce Eq.~\eqref{eq:disw} to Eq.~\eqref{eq:pi}. Therefore, they fail to make any correct prediction for Hamiltonians not respecting the inversion symmetry.  

To examine whether the neural networks have the ability to learn the winding number in its most general form, we generate training data with the most general one-dimensional Hamiltonians with chiral symmetry
\begin{equation}
H(k)=h_x(k)\sigma_x+h_y(k)\sigma_y, \label{eq:h}
\end{equation}
where $h_i(k)$, $i=x,y$ are periodic functions in $k$ expanded by the Fourier series
\begin{equation}
h_i(k)=\sum_{n=0}^c \left[a_{i,n} \cos(n k)+ b_{i,n}\sin(n k)\right]. \label{eq:fourier}
\end{equation}
$c$ is a cutoff that determines the highest possible winding number of the Hamiltonian, and is set to $ c=4 $ in the following. $a_{i,n}$, $b_{i,n}$ are randomly sampled from a uniform distribution within $ [-1,1] $. Among $10^5 $ training Hamiltonians, $37\%$, $50\%$ and $13\%$ of them having winding numbers $w=0$, $\pm 1$ and $\pm 2$, respectively. Different cutoff $ c $ and the distribution of $ w $ of the training data will not affect the network performance (See Supplemental Material). 

\begin{figure}[tbp]
	\includegraphics[width=.8\columnwidth]{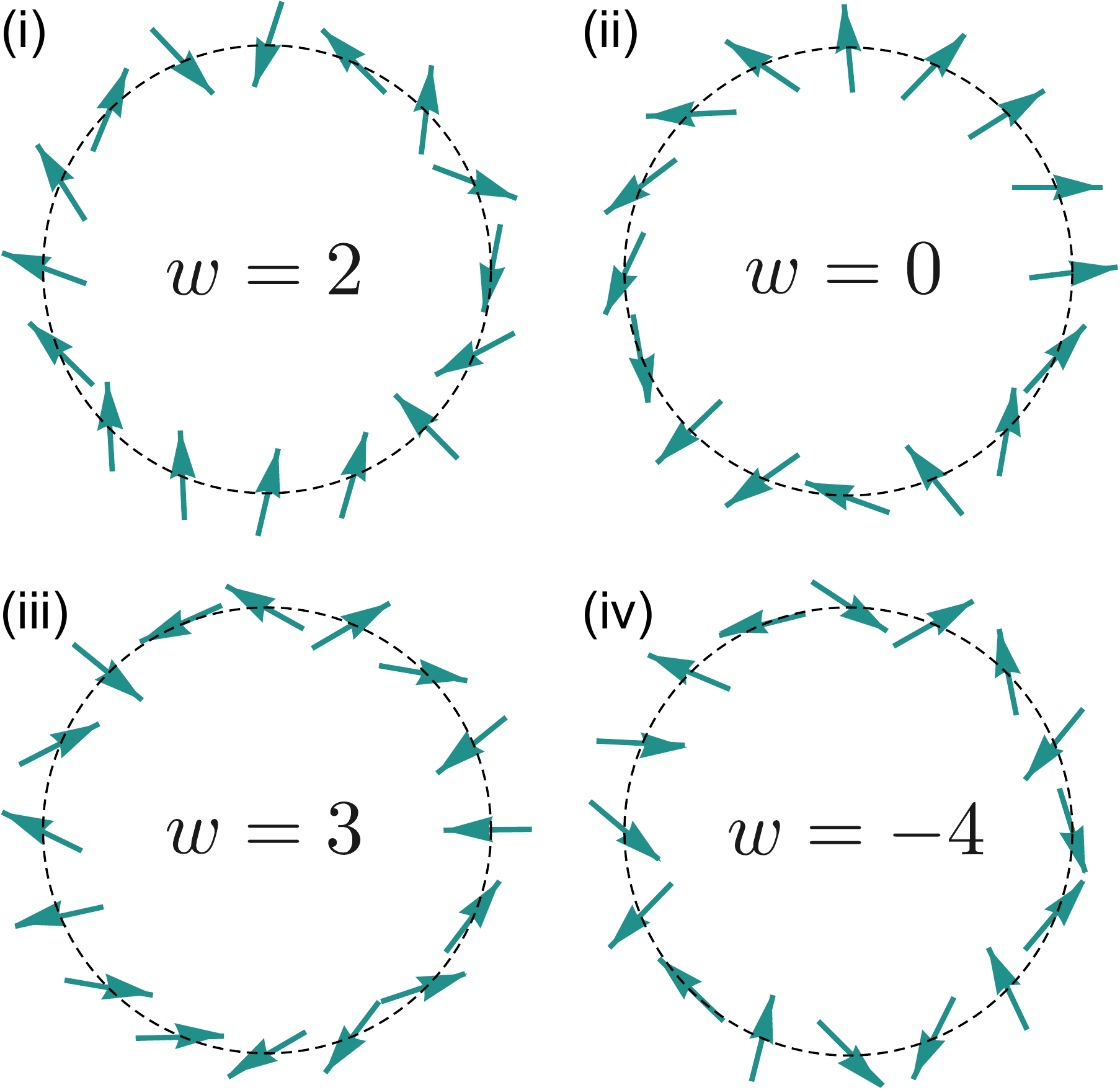}
	\caption{Typical Hamiltonians arranged in the Brillouin zone in the test set (i)-(iv). Here $ L=16 $. }
	\label{fig:configuration}
\end{figure}

We consider two classes of neural networks: the fully-connected network and the convolutional network \cite{Goodfellow2016book}. The fully-connected network has three hidden layers with 40, 32 and 2 neurons respectively. The total number of trainable parameters is 4061. The convolutional network has two convolutional layers with 40 kernels of size $ 2\times 2 $ and 1 kernel of size $ 1\times 1 $, followed by a fully-connected layer of 2 neurons before the output layer. The total number of trainable parameters is 310. The structure of the convolutional network is shown in Fig.~\ref{fig:structure}. All the hidden layers have rectified linear units $ f(x)=\max\{0,x\} $ as activation functions and the output layer has linear activation $ f(x)=x $.

We test these networks with four different test sets, schematically shown in Fig.~\ref{fig:configuration}. (i) $ 10^4 $ Hamiltonians with winding numbers $ w\in\{\pm2,\pm1,0\} $ that are not included in the training set; (ii) $ 10^4 $ Hamiltonians with the following functional form
\begin{equation}
\begin{split}
h_x(k)&=\theta(\pi-k)\cos f_1(k)\\
&\qquad\qquad+\theta(k-\pi)\cos[-f_2(k-\pi)+f_1(\pi)],\\
h_y(k)&=\theta(\pi-k)\sin f_1(k) \\
&\qquad\qquad+\theta(k-\pi)\sin[-f_2(k-\pi)+f_1(\pi)],
\end{split}
\label{eq:fool}
\end{equation}
where $ \theta(x) $ is the Heaviside step function, $ f_1(k) $ and $ f_2(k) $ are monotonic increasing functions bounded by $ f_1(0)=f_2(0)=0 $ and $ f_1(\pi)=f_2(\pi)\leq c\pi $. Intuitively, the Hamiltonian first winds the circle anticlockwisely during $ k\in[0,\pi] $, then clockwisely winds back during $ k\in[\pi,2\pi] $. The resulting winding numbers should always be zero; (iii) $ 10^4 $ Hamiltonians with winding numbers $ w=\pm 3 $; (iv) $ 10^4 $ Hamiltonians with winding numbers $ w=\pm 4 $. 

The test results are presented in Table.~\ref{tab:performance}. The convolutional network works generally better than the fully-connected network. The Hamiltonian configurations in Test (ii) have a strong local twist at $ k=\pi $ but the global topological numbers are always zero. That both neural networks perform well in this test is an indication that they have learned the global structures in the data instead of the local features. Surprisingly, the convolutional network can perform extremely well even in tests (iii) and (iv), which consist of Hamiltonians with larger winding numbers not seen by neural networks during the training. The fact that the convolutional network can pass test (iii) and (iv) shows that it has generalization power, and is also a strong indication that it really learns the general formula for the winding number. 

\begin{table}[tbp]
\centering
\caption{Performance (accuracy with respect to different test sets) of neural networks for learning topological phases in general models. }
\label{tab:performance}
\begin{tabular}{c|c|c|c|c}\hline
Network                   & Test (i) & Test (ii) & Test (iii) & Test (iv) \\\hline
Fully-connected & 82.2\% & 99.1\% & 22.8\% & 1.4\%   \\
Convolutional & 99.6\% & 100.0\%  & 98.2\% & 99.3\%   \\\hline
\end{tabular}
\end{table}

\textit{Open the black box.} Inspired by its performance, we open up the black box of the convolutional network and explore what it learns. Mathematically, our convolutional network can be described by the composition of the following functions:

(i) The first layer performs $ N=40 $ different convolutions with respect to the input Hamiltonians using the $ 2\times 2$ kernel $A^{i}$, $i=1,\ldots,N$: 
\begin{align}
\tilde{B}^i(n)&=A^i_{11}\tilde{h}_x(2\pi (n-1)/L)+A^i_{12}\tilde{h}_y(2\pi (n-1)/L) \notag\\
&+A^i_{21}\tilde{h}_x(2\pi n/L)+A^i_{22}\tilde{h}_y(2\pi n/L)+A^i_0,
\label{eq:first}
\end{align}
for $n=1,\ldots,L$, followed by $B^i(n)=f(\tilde{B}^i(n))$, where $f(x)$ is the activation function. 

(ii) The second layer performs another linear mapping across different kernels and is diagonal in $n$, i.e.
\begin{equation}
\tilde{D}(n)=\sum\limits_{i=1}^{N}c^i B^i(n)+c^0,
\label{eq:second}
\end{equation} 
followed by $D(n)=f(\tilde{D}(n))$.

(iii) Finally, the $L$-dimensional vector $D(n)$ is mapped to the winding number $\tilde{w}$ through 
\begin{align}
\tilde{F}_\eta=&\sum_{n=1}^L M_{\eta n}D(n)+N_\eta,\ \eta=1,2\\
F_\eta=&f(\tilde{F}_\eta),\\
\tilde{w}=&\sum_{\eta=1}^2 P_\eta F_\eta+Q.
\end{align}

All above, $A^i$, $c^i$, $M_{\eta n}$, $N_\eta$, $P_\eta$, and $Q$ are fitting parameters that are determined during the training. If the neural network successfully learns the discrete version of the winding number formula Eq.~\eqref{eq:disw}, we should expect $D(n)$ reproduces $\Delta\theta(n)$ and then the rest of the layers are basically summing over all $\Delta\theta(n)$. To verify this, we consider the input Hamiltonian
\begin{equation}
\begin{pmatrix}
\cos\phi & \cos(\phi+\Delta\phi) & \dots \\
\sin\phi & \sin(\phi+\Delta\phi) & \dots
\end{pmatrix}^T,
\label{eq:test}
\end{equation}
and expect 
\begin{equation}
D(n=1)\propto-\frac{1}{2\pi}\begin{cases}
\Delta\phi, &0\leq\phi<\pi, \\
\Delta\phi-2\pi, &\pi<\phi\leq 2\pi.
\end{cases}
\label{eq:dw}
\end{equation}

\begin{figure}[tbp]
	\includegraphics[width=0.95\columnwidth]{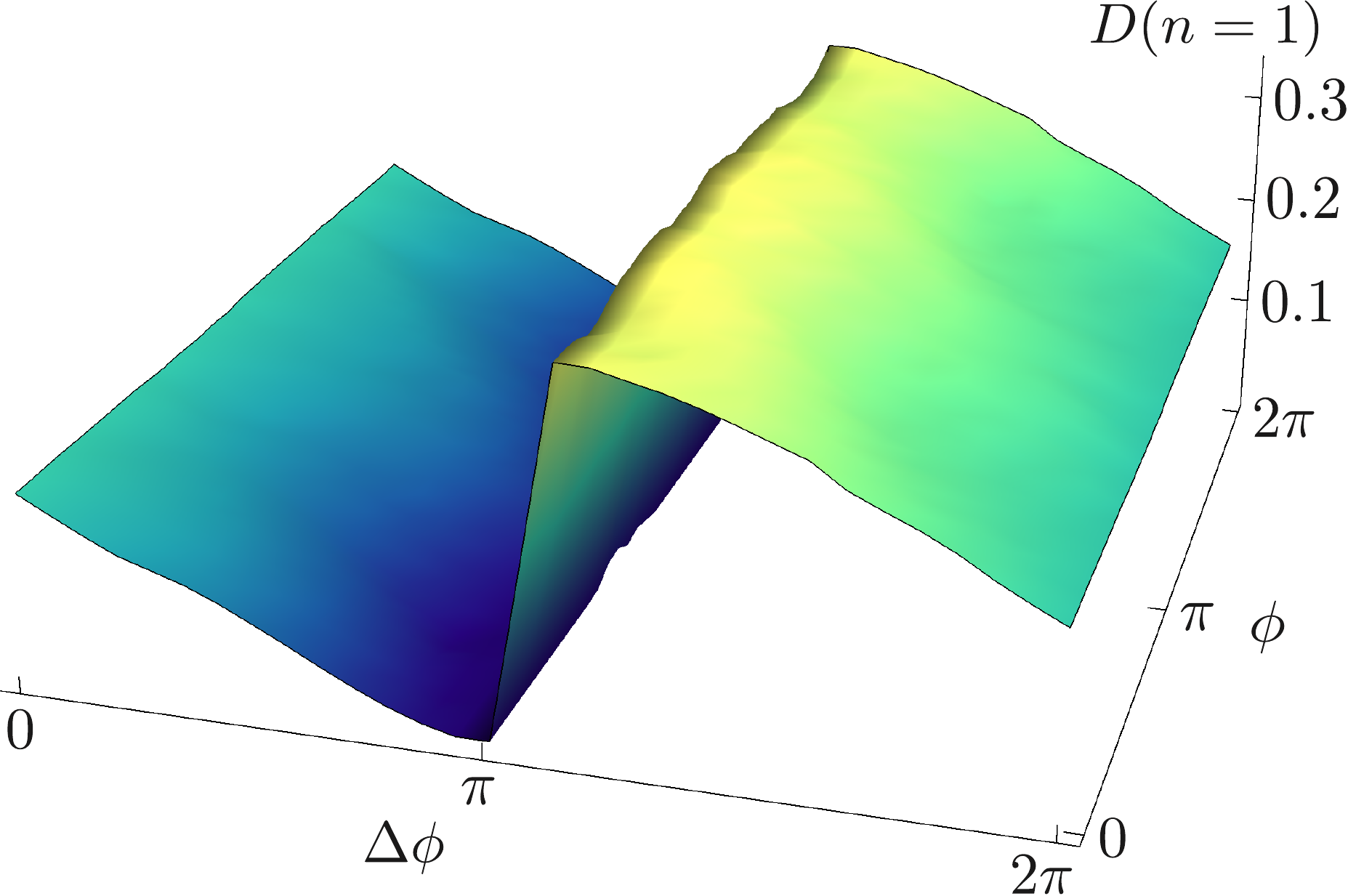}
	\caption{$ D(n=1) $ as a function of $\phi$ and $\Delta\phi$ when the input is Eq.~\eqref{eq:test}. }
	\label{fig:D}
\end{figure}

In Fig.~\ref{fig:D} we show $D(n=1)$ as a function of $\phi$ and $\Delta\phi$. It is very clear that, except for little fluctuations, $D(n=1)$ is independent of $\phi$ and depends on $\Delta\phi$ with the same function form as Eq.~\eqref{eq:dw}. 

With the above analysis, we can gain some understanding why our neural network has great generalization power. The convolutional layers that extract local windings are universal, and are unaffected by the global winding numbers of the data. As long as the training Hamiltonians are enough to cover the full surface of Fig.~\ref{fig:D}, the convolutional layers are always interpolating Eq.~\eqref{eq:dw} instead of extrapolating it, however large the global winding numbers are. Extrapolation only happens in the last two layers when $ \Delta\theta $ is summed. This is only a linear extrapolation, and is relatively easy for neural networks. In this way, the trained convolutional network computes winding numbers through the discrete version of the winding number formula Eq.~\eqref{eq:disw}.

\begin{figure}[tbp]
	\includegraphics[width=\columnwidth]{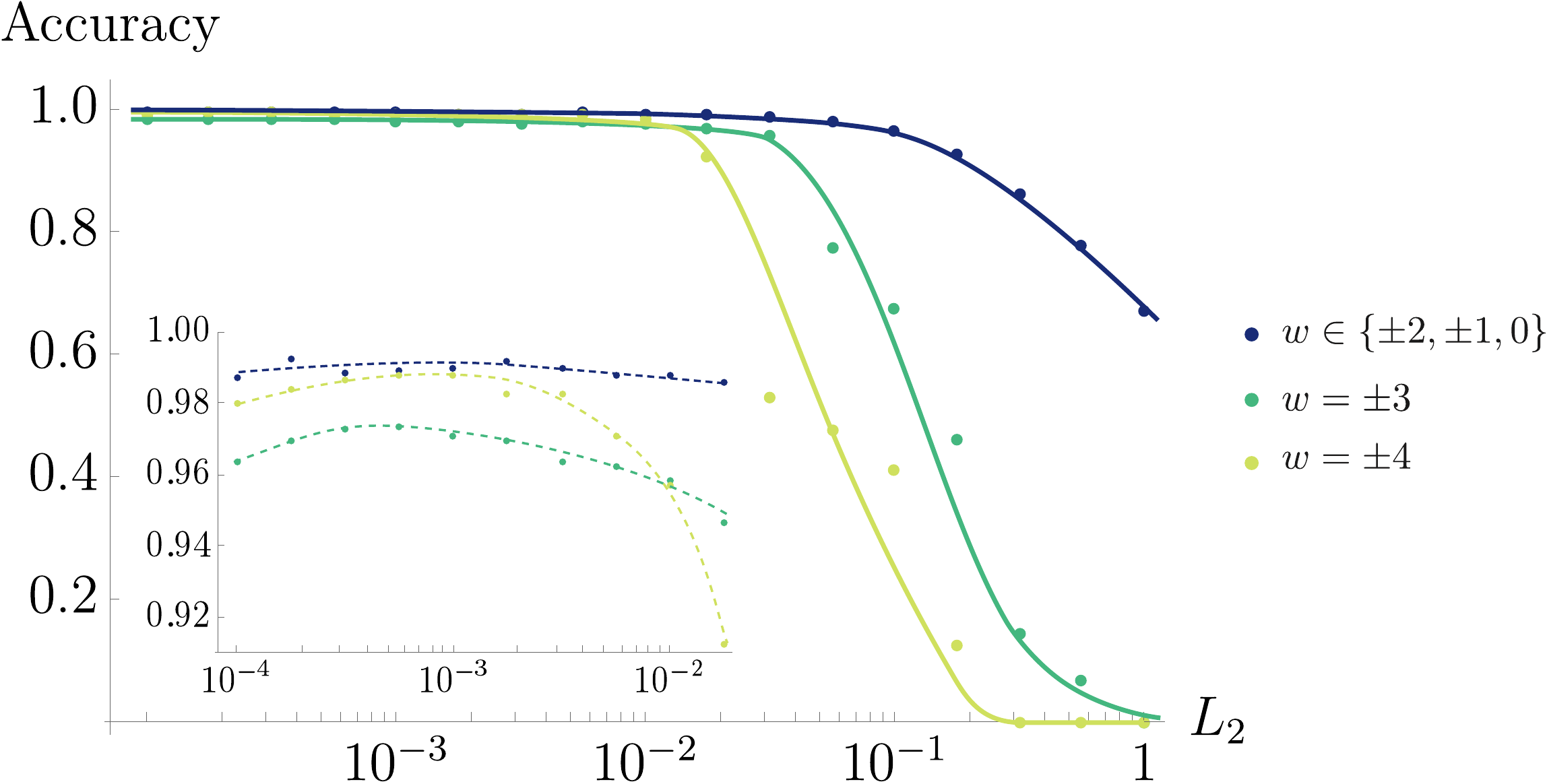}
	\caption{The performance of the convolutional network on various test sets with respect to the $ L_2 $ regularization strength. Solid lines, without introducing noise; inset dashed lines, randomly adding $ \pm 1 $ to the label of the winding number $ \omega $ for 4\% of the training data to mimic noise. }
	\label{fig:l2}
\end{figure}

\textit{Regularization techniques.} Finally, we remark on the regularization technique, which is usually considered necessary in training neural networks in order to avoid overfitting and to enhance networks' generalization power \cite{Bengio2012,Srivastava2014,Goodfellow2016book}. However, in our case we find the result to be contrary. In Fig.~\ref{fig:l2}, the ability of the network to compute larger winding numbers decays rapidly with the $ L_2 $ regularization strength, although the network could still very accurately compute winding numbers that are within the same range as the training set \footnote{All the results in this section are similar for other regularization techniques such as weight clipping. }. We attribute this phenomenon to the lack of noise. The data used here are generated by randomly sampling Hamiltonians \footnote{See also Ref.~\cite{VanNieuwenburg2017}. }, where there is much less noise, if noise exists at all. However, imagine the training data are taken directly from experiments. In this scenario the noise should exist and regularization should be useful. Indeed, this is demonstrated to be true in Fig.~\ref{fig:l2} if we artificially introduce noise into the training data. The situation is similar when data are generated by Monte Carlo sampling \cite{Carrasquilla2017,Broecker2017a,Chng2016,Ponte2017,Wang2016a,Liu2017c,Tanaka2017,Wetzel2017,Wetzel2017a,Hu2017a,Costa2017,Wang2017,Broecker2017,Chng2017}, where thermal noise may exist and regularization will be useful \footnote{Still, this depends on what to learn. Imagine the goal is to extract temperature from the Markov chain dynamics, these ``thermal noise'' is not noise but instead important features. }.

\textit{Concluding remarks.} In summary, we successfully train a neural network that learns global and nonlinear topological features from a large data set of Hamiltonians in the momentum space. We illustrate that our neural network has great generalization power to correctly compute larger winding numbers not seen in the training data. By analyzing the neural network, we confirm that our network does learn the discrete version of the winding number formula. Our network can directly be used to analyze the data from quantum simulators \cite{Flaschner2016}. It is also possible to generalize our results to the topological model in higher dimension and other classes \cite{NS}. We hope this work opens up a lot of possibilities of using machine learning to study rich topological physics.

Before concluding, we would like to make a couple of remarks on the role of symmetry when applying machine learning to physical problems. First, the symmetry of the training data matters. In order for neural networks to learn general rules, the training data have to be as general as possible to avoid unnecessary symmetry constraints. As demonstrated by the counterexample of learning the SSH model, the neural network exploits the inversion symmetry and learns a shortcut to the winding number. Second, the symmetry of the neural network matters. The structure of the neural network should be designed to be compatible with the symmetry of the targeting physics law. It is tempting to ask why the convolutional network performs better than the fully-connected network, as shown in Table.~\ref{tab:performance}, even though the later has more trainable parameters and hence greater expressibility in principle. This is because the translation of Hamiltonian configurations in the momentum space does not change the winding number. In practice, the translational symmetry is hard to be rediscovered for the fully-connected network during training. The convolutional network, on the other hand, respects this symmetry explicitly, reducing the redundancy in the parametrization. Thus, it is easier for the training algorithm to find the optimal fitting parameters. Furthermore, the winding number formula is the summation of many local phase winding $ \Delta\theta $. The convolutional layer takes this notion of locality directly through the $ 2\times 2 $ kernels. 
As a result, the convolutional network performs better than the fully-connected network. 

\textit{Acknowledgment.} This work is supported by MOST under Grant No. 2016YFA0301600 and NSFC Grant No. 11325418 and No. 11734010. H.S. thanks IASTU for hosting his visit to Beijing, where the key parts of this work were done. H.S. is supported by MIT Alumni Fellowship Fund For Physics.

\bibliography{Winding_Ref}

\widetext
\clearpage


\section*{Supplementary material}

# Supplemental Material for "Machine Learning Topological Invariants with Neural Networks"

Pengfei Zhang,[1] Huitao Shen,[2, *] and Hui Zhai[1, 3, †]

[1]*Institute for Advanced Study, Tsinghua University, Beijing, 100084, China*
[2]*Department of Physics, Massachusetts Institute of Technology, Cambridge, Massachusetts 02139, USA*
[3]*Collaborative Innovation Center of Quantum Matter, Beijing, 100084, China*

## I. NEURAL NETWORK DETAILS

*Network Design* Apart from the symmetry consideration mentioned in the main text that determines the general architecture of our network, here we briefly describe how do we determine the detailed structrue of the neural networks.

We start from a complicated multiple-layer convolutional network with a fully-connected layer before the output layer, which is standard architecture in computer vision. We then test its performance. As long as the performance is good, we keep reducing its complexity (number of layers, neurons, etc) until the performance drops down significantly. The resulting neural network is reported in the main text.

*Network Output Distribution* As explained in the main text, the output of the neural network is a real number $\tilde{w}$. The real winding number is interpreted as the integer that is closest to $\tilde{w}$. One may wonder if it is appropriate to say the network makes the correct prediction if the output is $\tilde{w} = 0.499$ while the real winding number is $w = 0$? Nevertheless, in our numerical experiments we found the networks tend to produce $\tilde{w}$ that are extremely close to integers. As shown in Fig. 1, we plot the distribution of $\tilde{w}$ from the convolutional network on different test data sets.

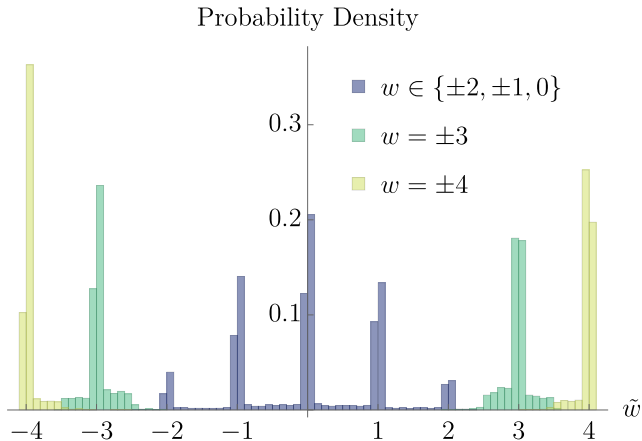

FIG. 1. (Color online) The distribution of output $\tilde{w}$ of the convolutional network on test data set (i), (iii) and (iv). The probability for a given test set (bins with the same color) sums to unity. There are narrow peaks at each integer.

## II. TRAINING DETAILS

*Training Algorithms* We use Adam algorithm [1] to minimize the mean squared error between $\tilde{w}$ and the output of the neural network and $w$. There is no regularization ($L_2$, dropout, etc) if not specified in the main text. We use mini-batch training with batch size to be 50. All the weights are randomly initialized to a normal distribution with Xavier method [2], and all the biases are initialized to zero. The networks are trained with Wolfram Mathematica's neural network module [3]. All other hyperparameters are set to be default without further explanation.

We confirm there is no overfitting in the training by always using a validation set. The validation set consists $10^4$ Hamiltonians with winding number $w \in \{0, \pm 1, \pm 2\}$ that are not in the training set. Typical loss during a training

---

* huitao@mit.edu
† hzhai@mail.tsinghua.edu.cn

instance of the convolutional network is shown in Fig. 2. Clearly there is no sign of overfitting. It takes about 100 epochs for the network to converge.

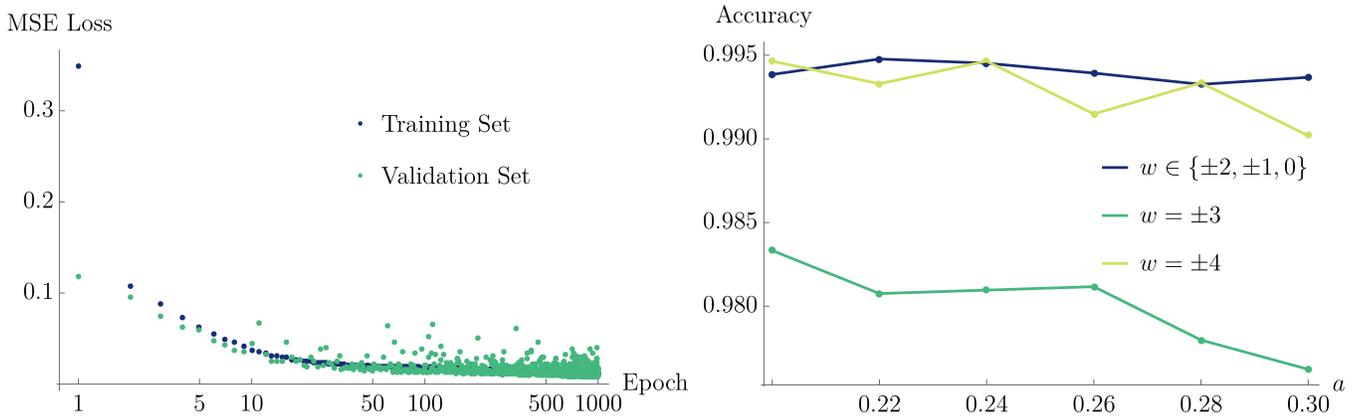

FIG. 2. (Color online) Left: Typical loss over different validation sets during a training instance of the convolutional network. Right: Performance of the convolutional network over test data sets (i), (iii) and (iv) with different training data distribution. The percentage of training data with $w = 0, \pm1$ and $\pm2$ are $a, 2a$ and $1 - 3a$.

*Training Data Distribution* As mentioned in the main text, the networks in the main text are trained using a training set of $10^5$ Hamiltonians, with 37%, 50% and 13% of which have winding numbers $0, \pm1$ and $\pm2$. Here we train the same network with different training data distributions: $a$, $2a$ and $1 - 3a$ of the total $10^5$ Hamiltonians have winding number $0$, $\pm1$ and $\pm2$, respectively. The result is shown in Fig. 2. It almost does not depend on the distribution.

In the main text, the cut off of the Hamiltonians for training data is $c = 4$ (See Eq. (7) in the main text) and we pick out Hamiltonians with $|w| > 2$ manually. Here we train the network with Hamiltonians of $c = 2$. 26%, 50% and 23% of their winding numbers are $0$, $\pm1$ and $\pm2$. The accuracies for the test set (i)-(iv) are 99.3%, 100%, 97.0% and 98.4% respectively. This is comparable with TABLE I in the main text.


\end{document}